\documentclass[10pt]{article}
\usepackage{amsmath}
\usepackage{amsfonts}
\usepackage{amssymb}
\usepackage{epsfig}
\usepackage{times}
\usepackage{oldgerm}
\newcommand{\be}{\begin{equation}}
\newcommand{\ee}{\end{equation}}
\newcommand{\ba}{\begin{array}}
\newcommand{\ea}{\end{array}}
\newcommand{\bea}{\begin{eqnarray}}
\newcommand{\eea}{\end{eqnarray}}

\newcommand{\rar}{\rightarrow}
\newcommand{\p}{\partial}
\newcommand{\ol}{\overline}
\newcommand{\ti}{\tilde}
\newcommand{\la}{\langle}
\newcommand{\ra}{\rangle}
\newcommand{\bct}{\begin{center}}
\newcommand{\ect}{\end{center}}

\renewcommand{\l}{\newline\null}
\def\figskip{\vskip .5cm plus 3mm minus 2mm}
\def\hbar{\not{\hbox{\kern-2.3pt $h$}}}
\def\psl{\not{\hbox{\kern-2.3pt $p$}}}
\def\Psl{\not{\hbox{\kern-2.3pt $P$}}}
\def\ksl{\not{\hbox{\kern-2.3pt $k$}}}
\def\qsl{\not{\hbox{\kern-2.3pt $q$}}}
\begin{document}
\begin{titlepage}
%
April 1999 \hfill PAR-LPTHE 99/12
\vskip 4cm
{\baselineskip 17pt
\bct
{\bf ASPECTS OF ELECTROWEAK PHYSICS FOR A COMPOSITE HIGGS BOSON;\break}
{\bf APPLICATION TO THE $\boldsymbol Z$ GAUGE BOSON DECAYING INTO\break
 TWO LEPTONS AND TWO PSEUDOSCALAR MESONS}
\ect
}
\vskip .5cm
\centerline{B. Machet
     \footnote[1]{Member of `Centre National de la Recherche Scientifique'}
     \footnote[2]{E-mail: machet@lpthe.jussieu.fr}
}
\vskip 5mm
\centerline{{\em Laboratoire de Physique Th\'eorique et Hautes \'Energies}
     \footnote[3]{LPTHE tour 16\,/\,1$^{er}\!$ \'etage,
          Universit\'e P. et M. Curie, BP 126, 4 place Jussieu,
          F-75252 PARIS CEDEX 05 (France)}
}
\centerline{\em Universit\'es Pierre et Marie Curie (Paris 6) et Denis
               Diderot (Paris 7)}
\centerline{\em Unit\'e associ\'ee au CNRS UMR 7589}
\vskip 1.5cm
{\bf Abstract:}  I study phenomenological aspects of the $SU(2)_L \times
U(1)$ electroweak physics of a Higgs boson transforming like a
quark-antiquark pair. A correspondence is established between its flavour
content, the hierarchy of quark condensates, and the leptonic decay constants
$f$ of pseudoscalar mesons;
the Higgs-mesons couplings  coming from the symmetry-breaking scalar
potential can then  be expressed in terms of the $f$'s, of the
Cabibbo-Kobayashi-Maskawa mixing angles, and of the mass of the Higgs boson.
Application is made to the decays of a $Z$ boson into two leptons and
two charged pseudoscalar mesons,  more specially $e^+e^-B^+ B^-$,
$e^+e^-D_s^+ D_s^-$ and $e^+e^-K^\pm \pi^\mp$; the last channel, involving new
types of flavour changing neutral currents in the scalar sector, is
characteristic of this approach.
Unlike in the standard model for quarks, the detection of two outgoing charged
$B$ mesons is hopeless, like, unfortunately, that of  $K^\pm \pi^\mp$;
the channel $D_s^+ D_s^-$ could have been missed in present experiments
for a Higgs mass lower than $65\,GeV$ because of insufficient  statistics.
For higher masses this Higgs becomes undetectable from such channels.
This shows that too drastic conclusions should not be drawn from an eventual
negative outcome from searches for the standard Higgs boson, that
simple variants of it are more elusive, may have escaped detection, and
could still escape in the near future.
\l
\smallskip

\vfill
\null\hfil\epsffile{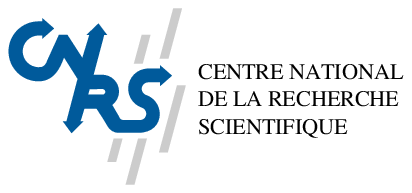}
\end{titlepage}
%
%
\section{Introduction}
\label{section:introduction}

The experimental search \cite{Higgsexp} for the Higgs \cite{Higgs}
boson is mainly concerned nowadays
by that of its strictly standard (unique) avatar \cite{GlashowSalamWeinberg}
or of its multiple supersymmetric \cite{SUSY} reincarnations.
The present and next generations of
accelerators should be able to give us precious information about this
cornerstone of particle physics.

If a positive outcome would undoubtedly be one of the greatest achievements
of the century and provide a fair reward for the simplest and most
elegant ideas,
a negative one should not be considered as a failure and be interpreted as
the absence of the Higgs
but rather as an encouragement to refine our ideas, to look for
alternatives, and in particular to investigate whether one
could have missed a Higgs particle with couplings smaller than
predicted in the basic models.

This is the goal of this work, which gives an example of a nearly standard
Higgs boson, still considered to be unique, which could have escaped
detection.

I outpass here the present negative feelings attached to technicolour
models \cite{technicolour}
(mainly due to the problem of flavour changing neutral currents)
and revive a composite Higgs boson; however, if it does transform like
a quark-antiquark pair, it nevertheless appears, unlike the quarks,
as a fundamental field in the Lagrangian, together with the observed
pseudoscalar (and scalar) mesons, along the lines of \cite{Machet1}.
It is a neat way of including both chiral and electroweak properties of
asymptotic states.
As the proposed model respects in particular the standard electroweak
transformations of quarks, the customary problems associated with flavour
changing neutral currents do not arise (we shall see that a new type of them
occurs but only in the Higgs sector, and beyond our present detection
capabilities).

The only concern being here the electroweak physics of scalar (Higgs) and
pseudoscalar  $J=0$ states, Quantum Chromodynamics (QCD) \cite{QCD}
and strong interactions are
deliberately left aside, except in that they determine the asymptotic
mesons and their normalization; interactions among particles in the final
state are not considered.  The quark picture is often invoked, but
only to make the reader more comfortable
and to provide a link with customary considerations: the quarks
themselves, which are not asymptotic fields, do not appear in the
Lagrangian.
The breakdown of the electroweak symmetry, tantamount to a condensation
of quark-antiquark pairs, is treated on a purely phenomenological
ground (much alike in the standard electroweak model) without reference to a
deeper underlying mechanism.

The Higgs boson being searched for through its decays,
determining its coupling to {\it observed} particles is essential.
Yukawa couplings to quarks are now absent, but the Higgs couples through a
``mexican hat'' potential to the three goldstones of the broken electroweak
symmetry, which are, as is shown by the study of their leptonic decays,
linear combinations of pseudoscalar mesons.
There exist $N^2/4$ ($N/2$ is the number of generations) real electroweak
quadruplets of scalar and
pseudoscalar (composite) fields isomorphic to the complex isoscalar doublet of
the Glashow-Salam-Weinberg model \cite{GlashowSalamWeinberg}.
Accordingly, the flavour orientation of
the Higgs boson is determined by a rotation in this $N^2/4$ dimensional space.
It determines, too, the flavour content of the three (pseudoscalar)
Goldstone bosons which are the three partners of the Higgs in the same
quadruplet, and the hierarchy between the  leptonic decay constants $f$ of
the various pseudoscalar mesons.
This makes possible the determination of the couplings of the Higgs to
pseudoscalar mesons as a function of the $f$'s, of Cabibbo-Kobayashi-Maskawa
(CKM) \cite{CKM} mixing angles and of the mass of the Higgs itself.

Its couplings to gauge fields keep as in the standard model.

I also show how, in an interpretation in terms of quarks, the orientation of
the Higgs boson determines the hierarchy of the
different quark-antiquark vacuum expectation values.

The paper is organized as follows:\l
- in section \ref{section:electroweak}, I recall the theoretical basis of the
model;\l
- in section \ref{section:EWbreaking}, I study the orientation of the
Higgs boson in flavour space and its consequences;\l
- in section \ref{section:decays}, I specialize to the three decays
$Z \rar e^+e^- B^+B^-$, $Z \rar e^+e^- D_s^+ D_s^-$ and $Z \rar e^+e^- K^\pm
\pi^\mp$;\l
- finally, appendix \ref{app:Dmatrix} gives technical details concerning
the basis of
electroweak and flavour eigenstates, and appendix \ref{app:GMOR} provides a
link between the
normalization of the asymptotic mesons and the Gell-Mann-Oakes-Renner
relation \cite{GellMannOakesRenner} in QCD.

\section{Electroweak interactions of quark-antiquark composite fields}
\label{section:electroweak}

\subsection{Electroweak and ``flavour'' eigenstates}
\label{subsec:eigenstates}

Though we only deal here with electroweak physics, the paper rests on
the fact that the ``asymptotic'' states for $J=0$ mesons can be interpreted
as ``flavour'' eigenstates, determined by strong interactions 
\footnote{Strong interactions are considered to be independent of
``flavour'', and so both have common eigenstates.}.

This can be put in correspondence with the existence of two different mass
scales, hence two different characteristic times:\l
- the electroweak mass scale $M_W \approx 80$ GeV, with an associated time
scale $\tau_{EW} = 1/M_W$;\l
- the mass scale associated with strong interactions, with an order of
magnitude of the masses of the mesons and resonances exchanged in nuclear
interactions, that is a few hundred MeV; the associated time scale
$\tau_S$ is much larger than $\tau_{EW}$; thus,  if an electroweak eigenstate
is produced (by electroweak interactions) at time $t$ it can only be detected
as such between $t$ and $t + \tau_{EW}$; after this interval, and before
it eventually decays into final states which can be non-hadronic, one only
detects its flavour components. The meaning of  ``asymptotic'', which has to
be adapted to the type of problem that is being analyzed, is
consequently here ``for time scales larger than $\tau_{EW}$''.

While the customary procedure is to try to incorporate strong interactions
between asymptotic electroweak eigenstates to build up observed particles,
which of course, like when introducing gluonic corrections in QCD,
faces non-perturbative problems, I will rather here
consider as perturbative and small the electroweak interactions between
flavour asymptotic states which are determined by strong interactions.
The only additional (non-perturbative) effect of strong interactions that
I will introduce is the normalization of asymptotic mesons which is
determined from their leptonic decays and is shown to be in agreement with
the ``Partially Conserved Current Hypothesis'' (PCAC) and with
the Gell-Mann-Oakes-Renner relation in QCD (see subsections
\ref{subsubsec:flavour}, \ref{subsubsec:norm}, appendix \ref{app:GMOR})
\footnote{
The question can be raised whether the Higgs
can appear as an asymptotic state or is projected at ``large'' times on
flavour (scalar) eigenstates;
its eventual detectability as an asymptotic particle can depend on it.
\label{foot:asympt}}
.
%
\subsection{Theoretical framework}
\label{subsec:theory}

The general framework has been set in \cite{Machet1}. For the sake of
understandability and for this paper to be self-contained, I briefly
recall here the main useful steps, in a somewhat less formal approach more
usable for phenomenological purposes.

Quarks are considered to be mathematical objects \cite{GellMann}
which are determined by their quantum numbers and by their transformations
by the different groups of symmetry that act upon them; we are mainly
concerned here with the chiral group $U(N)_L \times U(N)_R$ where $N$ is
the number of ``flavours'', and with the electroweak group
$SU(2)_L \times U(1)$: they form an $N$-vector $\Psi$
\begin{equation}
\Psi =
\left(
\ba{c}  u\\ c\\ \vdots \\d\\ s\\ \vdots \ea
\right)
\label{eq:Psi}\end{equation}
in the fundamental representation of the diagonal subgroup of the chiral group,
and their electroweak transformations, to which we stick to, are the usual
ones of the Glashow-Salam-Weinberg model \cite{GlashowSalamWeinberg}.

Any $SU(2)_L \times U(1)$ group can be considered, for $N$ even, as a subgroup
of $U(N)_L \times U(N)_R$; that it be the electroweak group, {\it i.e.}
that it act on quarks in the standard way determines its embedding
in the chiral group.
It is easy to check that the $SU(2)$ group the three generators of which are
the three $N \times N$ matrices (they are written in terms of four $N/2 \times
N/2$ sub-blocs with $\mathbb I$ the unit matrix and $\mathbb K$
the CKM mixing matrix)
\begin{equation}
{\mathbb T}^3_L = {1\over 2}\left(\begin{array}{rrr}
                        {\mathbb I} & \vline & 0\\
                        \hline
                        0 & \vline & -{\mathbb I}
\end{array}\right),\
{\mathbb T}^+_L =           \left(\begin{array}{ccc}
                        0 & \vline & {\mathbb K}\\
                        \hline
                        0 & \vline & 0           \end{array}\right),\
{\mathbb T}^-_L =           \left(\begin{array}{ccc}
                        0 & \vline & 0\\
                        \hline
                        {\mathbb K}^\dagger & \vline & 0
\end{array}\right),
\label{eq:SU2L}
\end{equation}
acting trivially on the left-handed projection $\Psi_L = [(1-\gamma_5)/2] \Psi$
of $\Psi$ can be identified with the standard electroweak $SU(2)_L$;
the $U(1)$ associated to the weak hypercharge $\mathbb Y$  is determined
through the Gell-Mann-Nishijima relation \cite{GellMannNijishima}
\begin{equation}
({\mathbb Y}_L,{\mathbb Y}_R) =
                       ({\mathbb Q}_L,{\mathbb Q}_R) - ({\mathbb T}^3_L,0),
\label{eq:GMN}\end{equation}
and from the trivial form for the (diagonal) charge operator $\mathbb Q$
\begin{equation}
{\mathbb Q}_L ={\mathbb Q}_R ={\mathbb Q}=\left(\begin{array}{ccc}
                        2/3 & \vline & 0\cr
                        \hline
                        0 & \vline & -1/3
           \end{array}\right).
\label{eq:Q}
\end{equation}
The $2N^2$ composite of the form $\bar q q$ or $\bar q\gamma_5 q$ can be cast
into $N^2/2$ quadruplets which are stable by the electroweak group; their
flavour structure is materialized by $N\times N$ matrices
$\mathbb M$ and the quadruplets can generically be written

\vbox{
\bea
& &\Phi(\mathbb D)=
({\mathbb M}\,^0, {\mathbb M}^3, {\mathbb M}^+, {\mathbb M}^-)(\mathbb D)\cr
& &\ \cr
& & =\left[
 \frac{1}{\sqrt{2}}\left(\begin{array}{ccc}
                     {\mathbb D} & \vline & 0\\
                     \hline
                     0 & \vline & {\mathbb K}^\dagger\,{\mathbb D}\,{\mathbb K}
                   \end{array}\right),
\frac{i}{\sqrt{2}} \left(\begin{array}{ccc}
                     {\mathbb D} & \vline & 0\\
                     \hline
                     0 & \vline & -{\mathbb K}^\dagger\,{\mathbb D}\,{\mathbb K}
                   \end{array}\right),
i\left(\begin{array}{ccc}
                     0 & \vline & {\mathbb D}\,{\mathbb K}\\
                     \hline
                     0 & \vline & 0           \end{array}\right),
i\left(\begin{array}{ccc}
                     0 & \vline & 0\\
                     \hline
                     {\mathbb K}^\dagger\,{\mathbb D} & \vline & 0
                    \end{array}\right)
             \right],\cr
& &
\label{eq:reps}
\eea
}
where $\mathbb D$ is a real $N/2 \times N/2$ matrix (see subsection
\ref{subsubsec:invariants} and Appendix \ref{app:Dmatrix}). One may
furthermore consider quadruplets the entries of which have a definite parity
(the $\mathbb S$'s below stand for scalars and the $\mathbb P$'s for
pseudoscalars)
\begin{equation}
\varphi = ({\mathbb S}^0, \vec {\mathbb P}),
\label{eq:SP}
\end{equation}
and
\begin{equation}
\chi = ({\mathbb P}\,^0, \vec {\mathbb S}).
\label{eq:PS}
\end{equation}
The $\varphi$'s and the $\chi$'s transform alike by the gauge group, according
to ($i$ and $j$ are $SU(2)$ indices)

\vbox{
\bea
{\mathbb T}^i_L\,.\,{\mathbb M}^j &=& -\frac{i}{2}\left(
              \epsilon_{ijk} {\mathbb M}^k +
                           \delta_{ij} {\mathbb M}^0
                              \right),\cr
{\mathbb T}^i_L\,.\,{\mathbb M}^0 &=&
                                \frac{i}{2}\; {\mathbb M}^i.
\label{eq:actioneven}
\eea
}
The link between the matrices $\mathbb M$ and diquark operators is 
straightforwardly established by sandwiching the latter between
$\ol\Psi$ and $\Psi$ and inserting a $\gamma_5$ when needed by  parity.

The link between diquark operators, of dimension $[mass]^3$, and the scalar
(pseudoscalar) fields (electroweak eigenstates) of dimension $[mass]$
occurring in the Lagrangian is achieved by introducing an appropriate
normalization as follows.

Let $H = h + \la H\ra$ be the Higgs boson such that
\begin{equation}
\la H \ra = \frac{v}{\sqrt{2}}
\label{eq:Hvev}
\end{equation}
breaks the electroweak $SU(2)_L \times U(1)$ into its electromagnetic
$U(1)_{em}$ subgroup.
Its flavour content is represented by an $N\times N$ matrix $\mathbb H$
and the associated diquark operator is $\ol\Psi {\mathbb H} \Psi$.
Eq.~(\ref{eq:Hvev}) is trivially satisfied by
\begin{equation}
H = \frac{\la H \ra}{\la \ol\Psi{\mathbb H}\Psi \ra} \ol\Psi {\mathbb H} \Psi
\label{eq:Hnorm}
\end{equation}
{from} which one can choose for all fields $\phi$ with dimension $[mass]$
associated with the $N\times N$ matrix $\mathbb M$ the normalization
\begin{equation}
\phi_{\mathbb M} = (i)\,\frac{\la H \ra}{\la \ol\Psi{\mathbb H}\Psi \ra}
                   \ol\Psi (\gamma_5){\mathbb M} \Psi.
\label{eq:phinorm}
\end{equation}
%

\subsubsection{Flavour eigenstates}
\label{subsubsec:flavour}

Because of the CKM rotation, the electroweak
eigenstates $\varphi$  and $\chi$ defined in (\ref{eq:SP},\ref{eq:PS})
are not flavour eigenstates but linear combinations of them.

The flavour or ``strong'' eigenstates are the ones associated with $\mathbb
M$ matrices which have only one nonvanishing entry equal to $1$.
Let ${\mathbb M}_{ab}$ such a matrix with one single non-vanishing entry at
the crossing of the $a$-th line and $b$-th column.
The associated flavour eigenstate, that we call $\Pi_{ab}$ is, according to
(\ref{eq:phinorm}), and in the case of a pseudoscalar
\begin{equation}
\Pi_{ab}({\mathbb M}_{ab}) = i\,\frac{\la H \ra}{\la \ol\Psi{\mathbb H}\Psi \ra}
                   \ol\Psi \gamma_5{\mathbb M}_{ab} \Psi.
\label{eq:flavoureigenstates}
\end{equation}
The $\Pi_{ab}$'s  are related to asymptotic states, {\it i.e.}
observed mesons ${\cal P}_{ab}$, by a scaling factor $\textswab b$
(see subsection \ref{subsubsec:norm}).

\subsubsection{Quadratic invariants and electroweak mass scales}
\label{subsubsec:invariants}

To every quadruplet $({\mathbb M}^0, \vec{\mathbb M})$
is associated a quadratic invariant:
\begin{equation}
{\cal I}
= ({\mathbb M}^0, \vec {\mathbb M})\otimes ({\mathbb M}^0, \vec {\mathbb M})
= {\mathbb {\mathbb M}}\,^0 \otimes {\mathbb {\mathbb M}}\,^0 +
                 \vec {\mathbb M} \otimes \vec {\mathbb M};
\label{eq:invar}
\end{equation}
the ``$\otimes$'' product is a tensor product (not the usual multiplication
of matrices) and means the product of fields as functions of space-time;
$\vec {\mathbb M} \otimes \vec {\mathbb M}$ stands for
$\sum_{i=1,2,3} {\mathbb M}\,^i \otimes  {\mathbb M}\,^i$.

For the relevant cases $N=2,4,6$, there exists a set of $\mathbb D$ real
matrices (see appendix \ref{app:Dmatrix}) such that the algebraic sum
of invariants specified below, extended over all  representations defined by
(\ref{eq:SP},\ref{eq:PS},\ref{eq:reps})
\begin{equation}
{1\over 2}
\left((\sum_{symmetric\ {\mathbb D}'s} - \sum_{antisym\ {\mathbb D}'s})
\left( ({\mathbb S}^0, \vec {\mathbb P})({\mathbb D})
                     \otimes  ({\mathbb S}^0, \vec {\mathbb P})({\mathbb D})
- ({\mathbb P}^0, \vec {\mathbb S})({\mathbb D})
                     \otimes  ({\mathbb P}^0, \vec {\mathbb S})({\mathbb D})
\right)\right)
\label{eq:diaginvar}
\end{equation}
is diagonal both in the electroweak basis and in the basis of flavour
eigenstates. With the coefficient $(1/2)$ chosen in (\ref{eq:diaginvar})
in the electroweak basis,
the normalization in the basis of flavour eigenstates is $(+1)$,
with all signs positive.

{From} the property stated above, to each quadruplet can be associated an
arbitrary electroweak mass scale and, for such a choice of
$\mathbb D$ matrices, the degeneracies of electroweak and flavour eigenstates
coincide.

\subsubsection{The electroweak Lagrangian}
\label{subsubsec:lagrangian}

The scalar (pseudoscalar) electroweak fields that build up the Lagrangian
are taken to be the ones associated with the set of matrices
$\mathbb D$ diagonalizing the invariant (\ref{eq:diaginvar}) in both the
electroweak and the flavour basis, and the combination used for the kinetic
terms is the one of (\ref{eq:diaginvar}).

The kinetic terms for the leptons and the gauge fields are the standard ones.

No Yukawa coupling to quarks is present since they are not fields of the
Lagrangian, and masses are given in a gauge invariant way to the mesons
themselves
\footnote{The hadronic sector being anomaly-free, a purely vectorial theory
in the leptonic sector is then favoured; the approach proposed in
\cite{BellonMachet} is an example of that, where leptons are
given masses without introducing Yukawa couplings and where the standard
model appears as an effective theory when one neutrino helicity decouples by
becoming infinitely massive.}
 .

A ``mexican hat'' potential is phenomenologically introduced to trigger the
spontaneous symmetry breaking of the electroweak symmetry.

\subsubsection{Normalizing the fields}
\label{subsubsec:norm}

By convention, we take all electroweak eigenstates $\phi$ occurring in the
Lagrangian and defined by (\ref{eq:phinorm}) to be normalized to ``$1$'':
\begin{equation}
\int \frac{d^4p}{(2\pi)^3}\vert\phi(p)\ra\la\phi(p)
\vert \theta(p_0) \delta(p^2 - m_\phi^2) = 1,
\label{eq:norm1}
\end{equation}
\begin{equation}
\la\phi(p')\vert\phi(p)\ra = 2p_0 (2\pi)^3 \delta^3(p-p'),
\label{eq:norm2}
\end{equation}
and, for fields corresponding to two different $\mathbb D$ matrices
\begin{equation}
\la\phi({\mathbb D}_\alpha)\vert \phi({\mathbb D}_\beta)\ra = 0\ ,
\ \alpha\not= \beta.
\label{eq:norm3}
\end{equation}
The phase-space measure for $\phi$ is
\begin{equation}
d\mu_{\phi(p)} =
\frac{d^4p}{(2\pi)^3} \theta(p_0) \delta(p^2 - m_\phi^2).
\label{eq:measure1}
\end{equation}
The flavour eigenstates $\Pi_{ab}$ defined in (\ref{eq:flavoureigenstates})
are then normalized to $(1/2)$
\begin{equation}
\la\Pi(p')\vert\Pi(p)\ra = (\frac{1}{2})2p_0 (2\pi)^3 \delta^3(p-p');
\label{eq:normPi}
\end{equation}
the difference of normalization between the $\phi$'s and the $\Pi$'s comes
from the identity
\begin{equation}
\sum_\alpha {\mathbb D}_\alpha \otimes {\mathbb D}_\alpha =
2 \sum_{ab} {\mathbb D}_{ab} \otimes {\mathbb D}_{ab}
\end{equation}
where the ${\mathbb D}_\alpha$'s are the ones exhibited in appendix
\ref{app:Dmatrix} and the ${\mathbb D}_{ab}$'s are $N/2 \times N/2$ matrices
the only non-vanishing entry of which is ``$1$'' at the crossing of the $a$-th
line and $b$-th column; this property also reflects into the
mismatch in the normalizations when expressing the quadratic invariant
(\ref{eq:diaginvar}) in terms of electroweak or flavour eigenstates, as
stated in subsection \ref{subsubsec:invariants}.

The last point concerns the normalization of the asymptotic (observed)
mesons ${\cal P}_{ab}$ (${\cal P}^+_{ud} =\pi^+, {\cal P}^-_{su} = K^-\ldots $)
which have the same flavour structure as the $\Pi_{ab}$'s:
one introduces the scaling factor $\textswab b$ such that  
\begin{equation}
\Pi = {\textswab b} {\cal P}.
\label{eq:bdef}
\end{equation}
For the ${\cal P}$'s one has
\begin{equation}
\la {\cal P}(p')\vert {\cal P}(p)\ra = \frac{1}{2\,{\textswab{b}^2}}2p_0
(2\pi)^3 \delta^3(p-p')
\label{eq:norm4}
\end{equation}
and
\begin{equation}
d\mu_{{\cal P}(p)} =
2\,{\textswab{b}^2}\frac{d^4p}{(2\pi)^3} \theta(p_0)
\delta(p^2 - m_{\cal P}^2).
\label{eq:measure2}
\end{equation}
The scaling factor $\textswab{b}$ is determined from the leptonic
decays of the pseudoscalar mesons and is
\footnote{In previous works \cite{Machet2} the asymptotic mesons were
normalized to ``$1$'', which led to a scaling factor
$a = 1/\textswab{b}$. Though the physical results turn out, as expected to be
the same, the procedure used here is more systematic and makes the links
more conspicuous with the traditional picture of QCD.}
\begin{equation}
\textswab{b} = \frac{\la H\ra}{2f_0}
\label{eq:b}
\end{equation}
where $f_0$ is the generic leptonic decay constant.

As a consequence, one has for example (for $N=4$)
\begin{equation}
P^+({\mathbb D}_1) =\textswab{b} \left(\cos\theta_c(\pi^+ + D_s^+) +
\sin\theta_c(K^+ - D^+)\right),
\label{eq:projection}
\end{equation}
where $\theta_c$ is the Cabibbo angle.

I show in Appendix \ref{app:GMOR} how this scaling is consistent,
through PCAC, with the Gell-Mann-Oakes-Renner relation
\cite{GellMannOakesRenner} in QCD.

The scaling factor $\textswab b$, that we introduce in a non-perturbative
way, and which ensures the consistency with other approaches and with
experimental observations, can be put in parallel with the $\sqrt{Z}$'s 
normalizing asymptotic fields in the $S$-matrix theory; our hypothesis lies
here in that, once admitted that strong interactions determine asymptotic
states, their only other effect can be phenomenologically parameterized
by $\textswab b$
\footnote{We in particular do not consider here the phase-shifts introduced
by strong interactions among final states.}.

\section{Electroweak symmetry breaking and the Higgs boson}
\label{section:EWbreaking}

\subsection{The  flavour orientation of the Higgs boson}
\label{subsec:Higgs}

The real $(H,\vec G)$ quadruplet (complex doublet) of the standard model,
where $H$ is the (scalar) Higgs boson  and $\vec G = (G^+, G^3, G^-)$ are
the three goldstones of the broken electroweak symmetry is isomorphic
to any of the $N^2/4$ quadruplets $\varphi$ of eq.~(\ref{eq:SP})
\footnote{
Considering that the Higgs boson is unique  prevents the occurrence of a
hierarchy problem \cite{GildenerWeinberg}}
.

We thus face an arbitrariness in the flavour content of the Higgs boson,
which also determines the composition of the three goldstones in terms of
flavour eigenstates since they are its three pseudoscalar partners in the
same quadruplet.

Identifying $H$ with ${\mathbb S}^0({\mathbb D}_1)$ as I did in previous
works \cite{Machet2} is tantamount to taking the same value for all diagonal
vacuum expectation values $\la \bar q_a q_a\ra$ ($a$ is a flavour index).
I loosen here this hypothesis and introduce a real orthogonal rotation
matrix $\cal{R}$ acting in the $N^2/4$ dimensional space of the
$\varphi$'s.

For the sake of simplicity, I will perform the analysis in the case of two
generations ($N=4$).
We can then restrict furthermore $\cal{R}$ to be a $3\times 3$ rotation
matrix by postulating that $\la {\mathbb S}^0({\mathbb D}_4)\ra =0$;
this is equivalent to saying that $\la \bar q_a q_b - \bar q_b q_a\ra =0$,
which is true if $CP$ is an unbroken symmetry.
So, the $(H,\vec G)$ multiplet is now considered to be a linear combination of
$\varphi_1=\varphi({\mathbb D}_1)$, $\varphi_2=\varphi({\mathbb D}_2)$ and
$\varphi_3=\varphi({\mathbb D}_3)$ (see eq.~(\ref{eq:SP})),
and the $4\times 4$ flavour matrix associated to $G^+$ reads
\begin{equation}
{\mathbb G}^+ = i\;\left(\begin{array}{ccc}
          0 & \vline & {\begin{array}{cc}G^+_{ud} & G^+_{us}\cr
                                         G^+_{cd} & G^+_{cs} \end{array}}
\\
                        \hline
                        0 & \vline & 0           \end{array}\right).
\label{eq:G+}
\end{equation}
Let the orthogonal matrix
\footnote{
The orthogonality of $\cal R$  preserves the property stated in subsection
\ref{subsubsec:invariants} that the quadratic combination
(\ref{eq:diaginvar}) is diagonal for both electroweak and flavour eigenstates;
suppose then that there is a single electroweak mass scale $M$ except
the vanishing one generated by the breaking of the electroweak symmetry;
since, would there be no goldstone, all flavour eigenstates
would have the same mass $M$, the ${\text(mass)}^2$ of the flavour components
of the charged goldstone $G^+$ are
\bea
M_{\pi^\pm}^2 &=& \frac{M^2}{2}(2- (G^+_{ud})^2),\cr
M_{K^\pm}^2 &=& \frac{M^2}{2}(2- (G^+_{us})^2),\cr
M_{D^\pm}^2 &=& \frac{M^2}{2}(2- (G^+_{cd})^2),\cr
M_{D_s^\pm}^2 &=& \frac{M^2}{2}(2- (G^+_{cs})^2).
\label{eq:masses}
\eea
Our statement that strong interactions determine asymptotic states for mesons
can thus participate to creating  a mass hierarchy for the latter as a
consequence of the breaking of the electroweak symmetry.
Of course, electroweak interactions among asymptotic states can occur,
in particular through the non-diagonal mass terms which do not cancel
any longer after the symmetry is broken.
}
$\cal R$ such that
\begin{equation}
\left( \begin{array}{c}
         \tilde\varphi_1 \\
         \tilde\varphi_2 \\
         \tilde\varphi_3 
       \end{array} \right)
\equiv \left( \begin{array}{c} (\ti{\mathbb S}_1, \vec{\ti{\mathbb P}_1}) \\ 
                          (\ti{\mathbb S}_2, \vec{\ti{\mathbb P}_2}) \\ 
                          (\ti{\mathbb S}_3, \vec{\ti{\mathbb P}_3})
       \end{array} \right)
= {\cal R} \left( \begin{array}{c} \varphi({\mathbb D}_1) \\
                                   \varphi({\mathbb D}_2) \\
                                   \varphi({\mathbb D}_3)
                  \end{array} \right)
\label{eq:rotation}
\end{equation}
depend on three mixing angles $\theta_1, \theta_2, \theta_3$,
the sines and cosines of which will be noted $s_1,s_2,s_3$ and $c_1,c_2,c_3$
\begin{equation}
{\cal R} = \left(    \begin{array}{ccc}
                 c_1       &        -s_1c_3        &       -s_1s_3    \\
                s_1c_2     &  c_1c_2c_3 - s_2s_3   &  c_1c_2s_3 + s_2c_3 \\
                s_1s_2     &  c_1s_2c_3 + c_2s_3   &  c_1s_2s_3 - c_2c_3
                     \end{array} \right).
\label{eq:R}
\end{equation}
We choose by convention 
\footnote{This makes in particular our results independent of $\theta_2$, in
relation with the fact that one relative phase between the quadruplets has no
physical significance.}
\begin{equation}
(H, \vec G) = \tilde\varphi_1 = (\ti{\mathbb S}_1, \vec{\ti{\mathbb P}_1}),
\label{eq:defHG}
\end{equation}
leading to
\bea
G^+_{ud} &=& c_\theta(c_1 - s_1s_3) + s_\theta s_1s_3, \\
G^+_{us} &=& s_\theta(c_1 - s_1s_3) - c_\theta s_1s_3, \\
G^+_{cd} &=& -s_\theta(c_1 + s_1s_3) - c_\theta s_1s_3, \\
G^+_{cs} &=& c_\theta(c_1 + s_1s_3) - s_\theta s_1s_3,
\label{eq:Gij}
\eea
where $s_\theta$ and $c_\theta$ stand for the sine and cosine of the Cabibbo
angle.

\subsection{Leptonic decays and the hierarchy of decay constants}
\label{subsec:leptonic}

I show here how the orientation of the Higgs boson in flavour space
determines the hierarchy of the pseudoscalar leptonic decay constants.

The leptonic decays are described by the diagram of Fig.~1; because the
non-diagonal $W-\text{meson}$ coupling only occurs for the three goldstones
$\vec G$, their flavour components are the only pseudoscalar mesons
which can decay leptonically.
These decays determine the scaling factor $\textswab b$
(\ref{eq:bdef},\ref{eq:b}) of the flavour eigenstates.

\vbox{
\figskip
\begin{center}
\epsfig{file=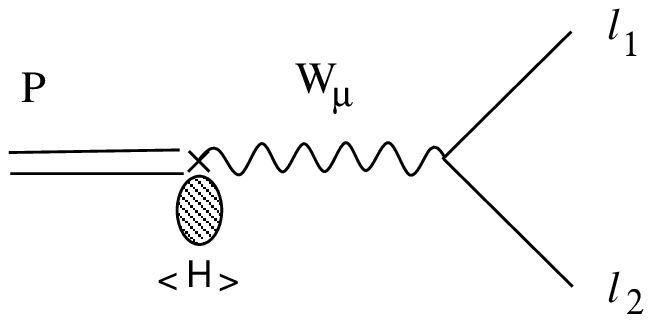}
\figskip
{\em Fig.~1: The leptonic decay of a pseudoscalar meson.}
\end{center}
\figskip
}

This section gives us the opportunity to explicitly compute a decay amplitude
involving asymptotic fields (pseudoscalar mesons) with a
normalization different from ``$1$'', while the leptons and gauge fields are
normalized to ``$1$''
\footnote{While the $\textswab b$ factor can be reabsorbed by a
simple rescaling of the mesonic fields when the Lagrangian involves only mesons,
like for example in the non-linear $\sigma$-model,
this is more problematic when several types of fields are involved which
interact between each other and the different kinetic terms of which come out
with different normalizations.}
.

The calculation proceeds by using a $W$ propagator in the unitary gauge,
and we do it here in analogy with the standard ``PCAC'' computation.

We have to evaluate
\begin{equation}
_{out}\la e^+ \nu_e \vert \pi^+\ra _{in}
 = \la e^+ \nu_e \vert \frac{G_F}{\sqrt{2}}\ L_\mu H^\mu\vert \pi^+\ra,
\label{eq:amplitude1}
\end{equation}
where $L_\mu$ and $H_\mu$ are respectively the weak leptonic and hadronic
currents.
$H_\mu$ is deduced from the part of the Lagrangian corresponding to the
$(H,\vec G)$ quadruplet
\begin{equation}
{\cal L} \ni -\frac{gv}{2\sqrt{2}}\left(
\partial^\mu G^+W_\mu^- + \partial^\mu G^-W_\mu^+ \right)+ \ldots
\label{eq:Lgoldstone}
\end{equation}
such that, by inserting the vacuum $\vert 0\ra\la 0 \vert$ in
(\ref{eq:amplitude1}) on gets
\footnote{This yields the exact result as if computed directly from the
diagram of Fig.~1.}
\begin{equation}
_{out}\la e^+ \nu_e \vert \pi^+\ra _{in}
 =\frac{G_F}{\sqrt{2}}\ \la e^+ \nu_e \vert L_\mu\vert 0\ra
      \la 0 \vert v\ \partial^\mu G^- \vert \pi^+ \ra.
\label{eq:amplitude2}
\end{equation}
In analogy with (\ref{eq:projection}), one has (still in the case of two
generations)
\begin{equation}
G^+ = {\textswab b}\ (G^+_{ud}\ \pi^+ + G^+_{us}\ K^+ + G^+_{cd}\ D^+
                            + G^+_{cs}\ D_s^+),
\label{eq:goldstone}
\end{equation}
and, from the relation
\begin{equation}
\pi^+ = \frac{1}{2{\textswab b}}\left(
          c_\theta(P^+({\mathbb D}_1) + P^+({\mathbb D}_2))
       -  s_\theta(P^+({\mathbb D}_3) + P^+({\mathbb D}_4)) \right)
\label{eq:pi+1}
\end{equation}
the inversion of the rotation (\ref{eq:rotation}) defining the
$(H,\vec G)$ multiplet (\ref{eq:defHG}) yields
\begin{equation}
\pi^+ = \frac{1}{2{\textswab b}}\left(
     (c_\theta (c_1 - s_1c_3) + s_\theta s_1s_3) G^+ + \ldots \right),
\label{eq:pi+2}
\end{equation}
and thus
\begin{equation}
_{out}\la e^+ \nu_e \vert \pi^+\ra _{in} =
\frac{1}{2{\textswab b}}
\left(c_\theta(c_1 - s_1c_3) + s_\theta s_1s_3)\right) \frac{G_F}{\sqrt{2}}\ 
\la e^+ \nu_e \vert L_\mu\vert 0\ra
\la 0 \vert v\ \partial^\mu G^- \vert G^+\ra;
\label{eq:amplitude4}
\end{equation}
as the goldstones have been normalized to ``$1$''
\begin{equation}
\la 0 \vert G^- \vert G^+\ra =1
\end{equation}
one obtains
\begin{equation}
_{out}\la e^+ \nu_e \vert \pi^+\ra _{in} =
i\, v\, k^\mu\;\frac{1}{2{\textswab b}}
\left(c_\theta (c_1 - s_1c_3) + s_\theta s_1s_3\right) \frac{G_F}{\sqrt{2}}\ 
\la e^+ \nu_e \vert L_\mu\vert 0\ra,
\label{eq:amplitude5}
\end{equation}
where $k^\mu$ is the momentum of the incoming pion.

When $\textswab b$ is given by (\ref{eq:b}), one recovers the standard PCAC
result for
\begin{equation}
f_\pi = \left((c_1 - s_1c_3) + \frac{s_\theta}{c_\theta}s_1s_3\right) f_0
      = \frac{G^+_{ud}}{c_\theta} f_0 .
\label{eq:fpi}
\end{equation}
It is easy to find along the same way
\bea
f_K &=& \left((c_1 - s_1c_3) - \frac{c_\theta}{s_\theta}s_1s_3\right) f_0
       = \frac{G^+_{us}}{s_\theta} f_0,\cr
f_D &=& \left((c_1 + s_1c_3) + \frac{c_\theta}{s_\theta}s_1s_3\right) f_0
       = \frac{G^+_{cd}}{(-s_\theta)} f_0,\cr
f_{D_s} &=& \left((c_1 + s_1c_3) - \frac{s_\theta}{c_\theta}s_1s_3\right) f_0
       = \frac{G^+_{cs}}{c_\theta} f_0.
\label{eq:f's}
\eea
In the limit $\theta_1 = \theta_3=0$ all $f$'s become identical.

Note the computation of the decay rate does not introduce extra
$\textswab b$ factors in the phase space integral  because the outgoing
states are leptons, which are normalized to ``$1$''.

\subsection{The hierarchy of quark condensates}
\label{subsec:condensates}

I show now in a precise example how the orientation of the Higgs in flavour
space also determines the hierarchy of quark condensates.

By our choice of a unique Higgs boson identified with the scalar entry of
$\tilde \varphi_1$, we have imposed that it is the only scalar with a
non-vanishing vacuum expectation value (VEV). This means a departure from the
symmetric case where all diagonal quark condensates have the same VEV and
where all non diagonal condensates vanish.

The system of equations to be satisfied by the different VEV's is now:\l
- for the scalar singlets of the $({\mathbb S}, \vec{\mathbb P})$
multiplets:
\begin{equation}
\la \ti{\mathbb S}_2\ra = \la \ti{\mathbb S}_3\ra =
\la{\mathbb S}^0({\mathbb D}_4)\ra = 0;
\end{equation}
- for the neutral scalars of the $({\mathbb P}^0, \vec{\mathbb S})$
multiplets:
\begin{equation}
\la {\mathbb S}^3({\mathbb D}_1)\ra = \la {\mathbb S}^3({\mathbb D}_2)\ra
= \la {\mathbb S}^3({\mathbb D}_3)\ra = \la {\mathbb S}^3({\mathbb D}_4)\ra
=0,
\end{equation}
where the notation $\la {\mathbb S} \ra$ used above is a shortcut for
$\la \ol\Psi {\mathbb S} \Psi\ra$.

One case for which these equations can be solved approximately is for
example $s_3 \approx 0, c_3 \approx 1$; one finds

\vbox{
\bea
& & \la \bar c c \ra \approx \frac{c_1 + s_1}{c_1 - s_1}\la \bar u u \ra, \cr
& & \la \bar d d \ra \approx \frac{c_1 -s_1(c_\theta^2 - s_\theta^2)}{c_1-s_1}
                    \la \bar u u \ra, \cr
& & \la \bar s s \ra \approx \frac{c_1 +s_1(c_\theta^2 - s_\theta^2)}{c_1-s_1}
                    \la \bar u u \ra, \cr
& & \la \bar d s \ra = \la \bar s d \ra \approx  -2 \frac{s_1 s_\theta
                c_\theta}{c_1 - s_1} \la \bar u u\ra, \cr
& & \la \bar u c \ra = \la \bar c u \ra = 0.
\label{eq:consensates}
\eea
}

For $c_1>0\;,\;s_1<0$, one gets accordingly the hierarchy
\begin{equation}
\la \bar u u \ra > \la \bar d d \ra >
\la \bar s s \ra >\la \bar c c \ra >\la \bar d s \ra =\la \bar s d \ra >
\la \bar u c \ra = \la \bar c u \ra = 0;
\end{equation}
it agrees with the one generally admitted from the Gell-Mann-Oakes-Renner
relation; the non-vanishing of non-diagonal quark condensates,
which appears in a natural manner here, has been debated in the
past when dealing with kaon decays \cite{tadpole}.
\subsection{The coupling of the Higgs boson to pseudoscalar mesons}
\label{subsec:higgs-mesons}

I determine here the coupling of the Higgs boson to pseudoscalar mesons
in terms of leptonic decay constants and CKM mixing angles.

The ``mexican hat'' potential introduced for $\ti\varphi_1 = (H,\vec G)$
\begin{equation}
V(H,\vec G)= -\frac{\sigma^2}{2} \ti\varphi_1\otimes\ti\varphi_1
          + \frac{\lambda}{4} (\ti\varphi_1\otimes\ti\varphi_1)^{\otimes 2}
\label{eq:potential}
\end{equation}
which triggers the breaking of the electroweak
symmetry yields in particular a coupling between the Higgs and the goldstones
\begin{equation}
{\cal L} \ni  -\frac{\lambda}{\sqrt{2}}v\; h\; (2G^+G^- + G^3 G^3).
\label{eq:phi4}
\end{equation}
Let $a,b$ the flavour indices of a pseudoscalar meson ${\cal P}_{ab}$,
with mass $M_{ab}$
transforming like the diquark operator $\bar q_a \gamma_5 q_b$ (for example
${\cal P}^+_{ud} = \pi^+$);
the goldstone $G^+$ writes, according to
(\ref{eq:phinorm},\ref{eq:flavoureigenstates},\ref{eq:bdef}) and in analogy
with (\ref{eq:goldstone})
\begin{equation}
G^+ = \sum_{a,b} G^+_{ab} \Pi^+_{ab}
    =  {\textswab b} \sum_{a,b} G^+_{ab} {\cal P}^+_{ab}\;.
\end{equation}
Making use of the equivalent of relations (\ref{eq:fpi},\ref{eq:f's}) in the
case of three generations, one gets in
particular for two charged outgoing flavour eigenstates ${\Pi}^+_{ab}$
and ${\Pi}^-_{cd}$ the coupling
\footnote{The Lagrangian is always written with fields which are normalized
to ``$1$'', and the corresponding couplings are used to compute $S$-matrix
elements between states which are also normalized to ``$1$''. The physical
$S$-matrix elements involving asymptotic mesons that we need to compute are
then deduced by introducing the appropriate $\textswab b$ factors.}
\begin{equation}
-i\, \sqrt{2}\lambda\,v\;\sum_{a,b}\sum_{c,d}
\frac{f_{ab}f_{cd}}{f_0^2}V_{ab}V^\dagger_{cd}
\;h\,{\Pi}^+_{ab}\,{\Pi}^-_{cd}
\label{eq:h2mesons}
\end{equation}
where $f_{ab}$ is the leptonic decay constant of ${\cal P}_{ab}$
and $V_{ab}$ the corresponding entry of the CKM mixing matrix.

A dominant feature in (\ref{eq:h2mesons}) is the presence of the CKM
mixing angles $V_{ab}$, which, unlike what happens at the quark level for
Yukawa couplings, can strongly damp the corresponding coupling independently
of the mass of the outgoing particles. This is specially the case for
$B^\pm$ mesons, since the corresponding $V_{ub}$ lies far from the diagonal
in the CKM mixing matrix.

I now use the above results to investigate decays of the $Z$ boson which are
mediated by the Higgs.

\section{Some decays
$\boldsymbol{ Z \rightarrow e^+e^-{\cal P}_i^+{\cal P}_j^-}$}
\label{section:decays}

I shall always deal with the cases when the outgoing leptons are electrons;
when they are muons, the results are very much alike, because the relative
difference in the available phase space is  very small.

\subsection{The decay $\boldsymbol{Z \rightarrow e^+e^-\,B^+B^-}$}
\label{subsec:BB}

There are two types of contributions shown in Figs.~2a,2b.

\vbox{
\bct
\epsfig{file=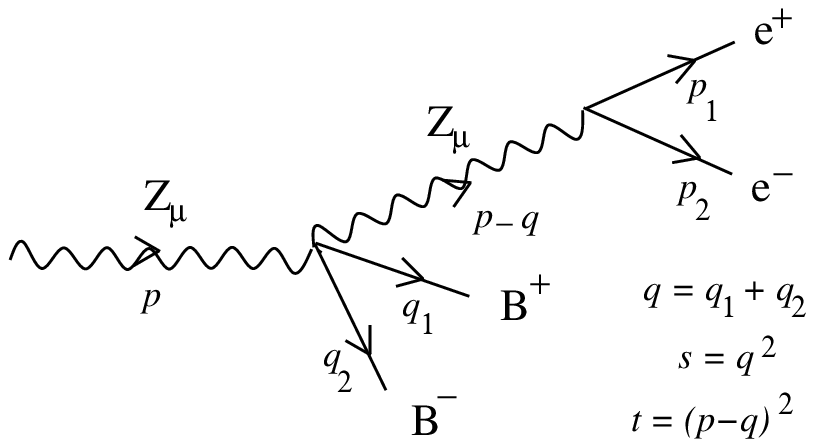,height=6truecm,width=8truecm}
\figskip
{\em Fig.~2a: The decay $Z \rar e^+e^- B^+B^-$: direct coupling.}
\ect
}

\vbox{
\bct
\epsfig{file=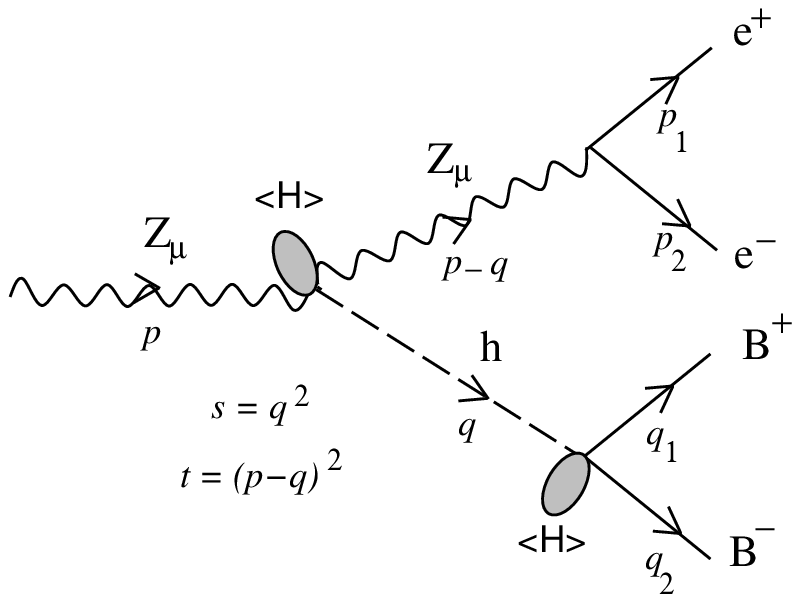,height=6truecm,width=8truecm}
\figskip
{\em Fig.~2b: The decay $Z \rar e^+e^- B^+B^-$: the contribution of the Higgs
boson.}
\ect
}

That there exists a
direct coupling of two gauge bosons to two mesons differs from the standard
model with quarks, where the mesons can only originate from the
``hadronization'' of the two quarks coupled to the Higgs boson trough a
Yukawa coupling.

We need to compute the amplitude ${\cal A}_{Z \rar B^+B^-e^+e^-}$
\begin{equation}
{\cal A}_{Z \rar B^+B^-e^+e^-} = {_{out}\la e^+e^-\,B^+B^-\vert Z_\mu\ra _{in}}
\label{eqq:amp1}
\end{equation}
where the asymptotic fields $Z_\mu, e^\pm$ are normalized to ``$1$'' while the
$B$ mesons are not (see subsection \ref{subsubsec:flavour}).
This introduces $\textswab b$ factors according to
\begin{equation}
{\cal A}_{Z \rar B^+B^-e^+e^-} = \frac{1}{{\textswab b}^2}\;
{_{out}\la e^+e^-\,\Pi^+_{ub}\Pi^-_{bu}\vert Z_\mu\ra _{in}}
\label{eq:amp2}
\end{equation}
where we have introduced the flavour eigenstates $\Pi_{ab}$
normalized to ``$1$'' which are also used to express the Lagrangian.
Eq.~(\ref{eq:amp2}) can be calculated with standard rules.

The total  coupling of two $Z$ bosons to $\Pi^+_{ub}\Pi^-_{bu}$ reads
(including the ``$i$'' coming from $\exp(iS)$)
\begin{equation}
i\,\frac{g^2}{2c_W^2}\left( 1 +
  \frac{1}{2}(\frac{f_B}{f_0})^2 V_{ub}V^\dagger_{bu}\frac{M_H^2}{q^2 - M_H^2}
                     \right)
\label{eq:ZPP}
\end{equation}
where the first contribution comes from the direct coupling and the second
from the one involving the Higgs boson; $c_W$ is the cosine of the Weinberg
angle; $q$ is the Higgs momentum, the mass of which is (see
(\ref{eq:potential},\ref{eq:Hvev}))
\begin{equation}
M_H^2 = \lambda v^2.
\label{eq:MH}
\end{equation}
Note that the mixing angles $V_{ub}V^\dagger_{bu}$ only appear in the Higgs
contribution.

A second potential  source of damping appears in (\ref{eq:ZPP}) since there
can be a destructive interference between the direct contribution and the
one with the Higgs boson.

The decay rate is expressed as a double integral over the square of the Higgs
momentum $s = q^2$ and that of the virtual $Z$ momentum $t = (p-q)^2$
:

\vbox{
\bea
\Gamma_{Z\rar e^+e^-B^+B^-}  &=&
\frac{1}{512\sqrt{2}\pi^5}(1-2s_W^2)M_Z^3 G_F^3\cr
& &\int_{4M_B^2}^{(M_Z - 2m_e)^2}ds\ 
\frac{\lambda^{1/2}(s,M_B^2,M_B^2)}{s}
      \left( 1 + \frac{1}{2}(\frac{f_B}{f_0})^2 V_{ub}V^\dagger_{bu}
                \frac{M_H^2}{s - M_H^2} \right)^2\cr
& &\int_{4m_e^2}^{(M_Z - \sqrt{s})^2}dt\ 
\frac{\lambda^{1/2}(M_Z^2,s,t)} {(t-M_Z^2)^2}
\left( t +\frac{\lambda(M_Z^2,s,t)}{12\,M_Z^2}\right),
\label{eq:gammaBB}
\eea
}

where $M_Z$ is the mass of the $Z$ gauge boson,
\footnote{With the normalizations chosen here one has
$M_W^2 = g^2v^2/8$ and $G_F/\sqrt{2} = g^2/8M_W^2 = 1/v^2$, where $g$ is the
$SU(2)_L$ coupling constant.}
$M_B$ the mass of the $B^\pm$ pseudoscalar mesons, $m_e$ the mass
of the electron
\footnote{The mass of the electron has been neglected in the computation of
the square of the amplitude.}
and $s_W$ the sine of the Weinberg angle;
$\lambda(u,v,w)$ is the fully symmetric function
\begin{equation}
\lambda(u,v,w) = u^2 + v^2 + w^2 -2\,uv -2\,uw -2\,vw;
\label{eq:lambda}
\end{equation}
All $\textswab b$ factors cancel in the decay rate:
the $1/{\textswab b}^4$ coming from the amplitude (\ref{eq:amp2}) squared
exactly matches the two factors ${\textswab b}^2$ coming from the
phase-space measures (\ref{eq:measure2}) for the two outgoing mesons.

The kinematical intervals being respectively $s \in [4M_B^2, (M_Z - 2m_e)^2]$
and $t \in [4m_e^2, (M_Z - \sqrt{s})^2]$,
for $M_H^2 < (M_Z - 2m_e)^2$, the potential divergence of the decay rate due to
the (squared) propagator of the Higgs has to be smoothed out by introducing
a width for the latter;
we restrict ourselves to the two-mesons channels, as depicted in Fig.~3.

\vbox{
\figskip
\bct
\epsfig{file=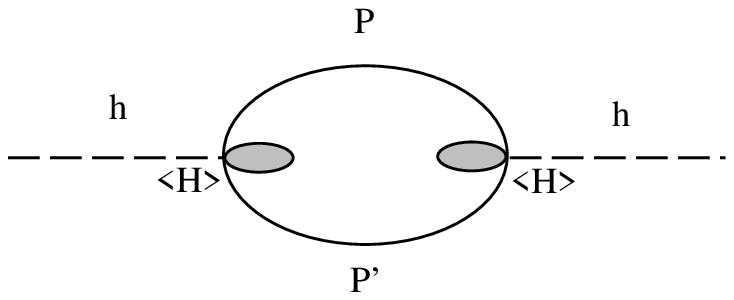,height=4truecm,width=8truecm}
\figskip
{\em Fig.~3: Introducing a width for the Higgs boson.}
\ect
}

The couplings of the Higgs to two charged mesons are the same as the ones
used above and studied in subsection \ref{subsec:higgs-mesons};
we shall however neglect in the computation of its width the misalignment
of the Higgs with respect to ${\mathbb S}^0({\mathbb D}_1)$, consider all
ratios $f_{ab}/f_0 \approx 1$ and only take into account the channels
which are not damped by small mixing angles (combinations of
$\pi^\pm$ and $D_s^\pm$ in the charged sector).
The neutral mesons ($\pi^0, \eta, \eta_c, \eta_b$) are incorporated along
the same lines.

We thus replace in (\ref{eq:gammaBB})

\vbox{
\bea
& &\left(
1 + \frac{1}{2}(\frac{f_B}{f_0})^2 V_{ub}V^\dagger_{bu}\frac{M_H^2}{(s-M_H^2)}
\right)^2
                        \longrightarrow \cr
&=& 1 + 
\frac
{(\frac{f_B}{f_0})^2 V_{ub}V^\dagger_{bu}M_H^2(s-M_H^2) +
\left( \frac{1}{2}(\frac{f_B}{f_0})^2 V_{ub}V^\dagger_{bu}\right)^2 M_H^4}
{(s-M_H^2)^2 + \left(\frac{G_FM_H^4}{32\pi\sqrt{2}}\right)^2
          \left(\sum_{ab}\sum_{cd} V_{ab}V^\dagger_{ba} V_{cd}V^\dagger_{dc}
        \frac{\lambda^{1/2}(s,M_{ab}^2,M_{cd}^2)}{s}\right)^2}\cr
& &
\label{eq:largeur}
\eea
}
where the $\sum_{ab} \sum_{cd}$ in (\ref{eq:largeur}) are performed over all
above mentioned couples of $J=0$ pseudoscalar mesons with flavour indices
$(ab)$ and $(cd)$ and masses $M_{ab}$ and $M_{cd}$;
in the charged sector, the $V_{ab}$'s are restricted to the diagonal entries
of the CKM mixing matrix, and stand for $1$ in the neutral
sector (see (\ref{eq:reps}) with ${\mathbb D} = 1$).

To maximize the decay rate, we take $f_B/f_0 \approx 10$ and
$V_{ub}$ is chosen at the upper limit of the experimental bounds
\cite{TableParticleProperties}
\begin{equation}
V_{ub} \approx 4\,10^{-3}.
\label{eq:Vub}
\end{equation}

\vbox{
\figskip
\bct
\epsfig{file=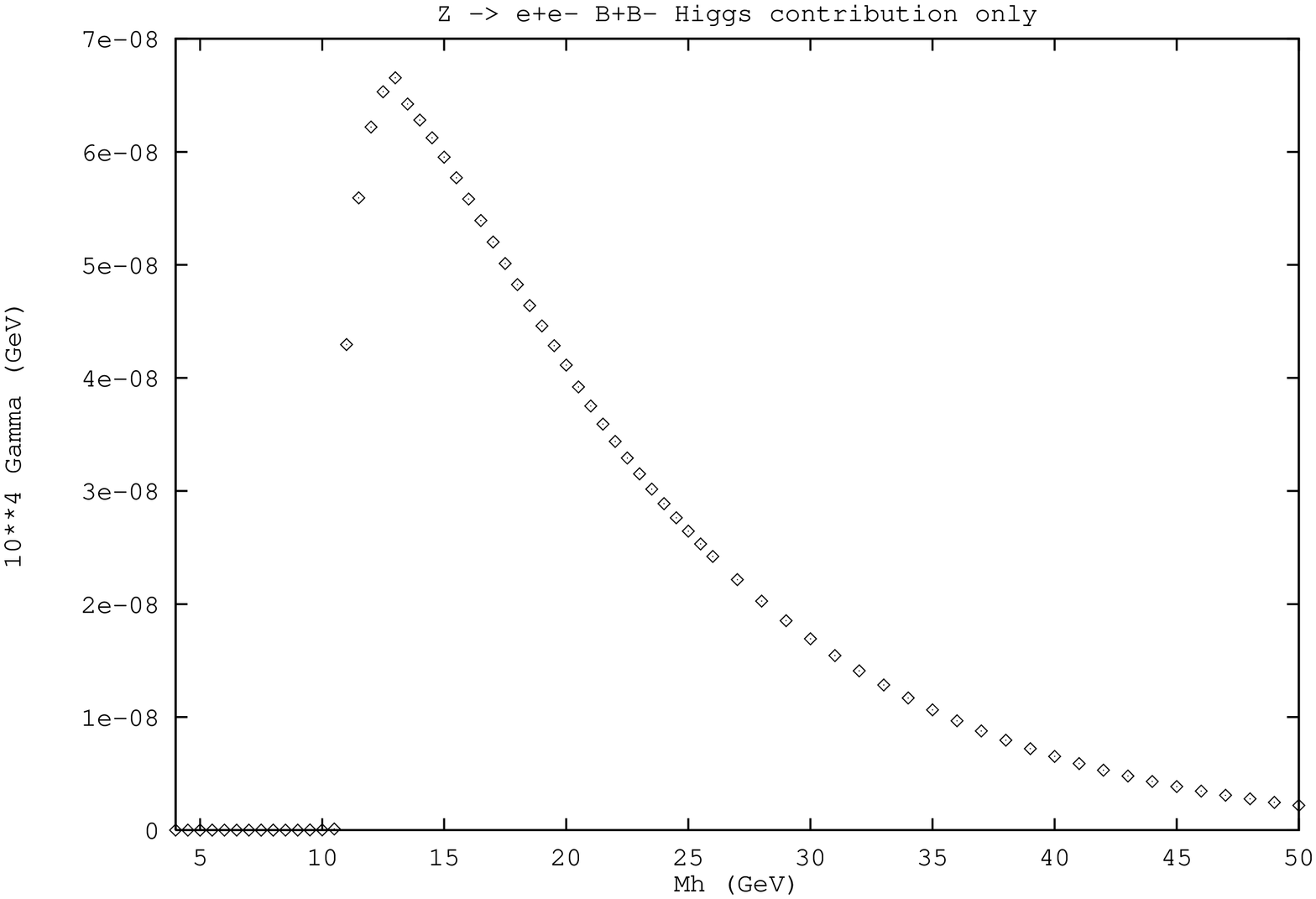,height=14truecm,width=10truecm,angle=-90}
\figskip
{\em Fig.~4: $10^4 \times$ the Higgs contribution to the decay rate
$Z \rar e^+e^- B^+B^-$.}
\figskip
\ect
}

The Higgs contribution, shown in Fig.~4, exhibits a resonance-like
behaviour, with a maximum
of $\approx 6.6\,10^{-12}\,GeV$ for $M_H \approx 13\;GeV$, which is
extremely small
\footnote{The numerical integrations have been performed by the method of
Newton and dividing, for each point of the graphs, the
two dimensional domain of integration  into $5\,10^3 \times 10^3$ cells,
ensuring a perfect stability.}
.
The ``background'' coming from the contribution
where no Higgs is involved is nearly $1000$ times larger than the Higgs
contribution at its maximum and is itself outside the reach of present
observations, with a rate
\footnote{The destructive interference effects between the two contributions
is of course negligeable.}
\begin{equation}
\Gamma^{no\ Higgs}_{Z\rightarrow e^+e^- B^+B^-} = 5\,10^{-9}\;GeV,
\label{eq:BBnoH}
\end{equation}
to be compared with the total width of the $Z$ boson
$\Gamma_Z \approx 2.4\;GeV$; suppose that one can analyze $20\, 10^6$ $Z$
decays \cite{EWmeasures}; ten identified decays into two leptons and two
pseudoscalar mesons would correspond to a
fraction $5\,10^{-7}$ of all and to a partial width
$\Gamma_{part} = 1.2\,10^{-6}\, GeV$; it seems consequently reasonable to
set  an (optimistic) threshold of observability above a partial width 
\begin{equation}
\Gamma_{obs} \geq 10^{-6}\,GeV.
\label{eq:thresh}
\end{equation}

Accordingly, it appears useless to look for a  Higgs boson like the one
described above in the decay $Z \rar e^+e^- B^+B^-$.

\subsection{The decay $\boldsymbol{Z \rightarrow e^+e^-\,D_s^+D_s^-}$}
\label{subsec:FF}

I then analyze the decay $Z \rightarrow e^+e^- D_s^+D_s^-$;
the corresponding rate is computed from formulae similar to
(\ref{eq:gammaBB},\ref{eq:largeur}), with the appropriate substitutions
concerning the masses, leptonic decay constants and mixing angles.
The damping due to CKM mixing angles is now negligeable, which makes the Higgs
contribution dominate over the direct coupling of the $Z$ to two mesons.
It appears consequently as a better reaction to look for the Higgs boson.

The background coming from from the direct coupling of the $Z$ to two mesons
is negligeable:
\begin{equation}
\Gamma^{no\ Higgs}_{Z\rightarrow e^+e^- D_s^+D_s^-} = 7.6\,10^{-9}\;GeV.
\label{eq:FFnoH}
\end{equation}
The results are shown in Figs.~5.

\vbox{
\figskip
\bct
\epsfig{file=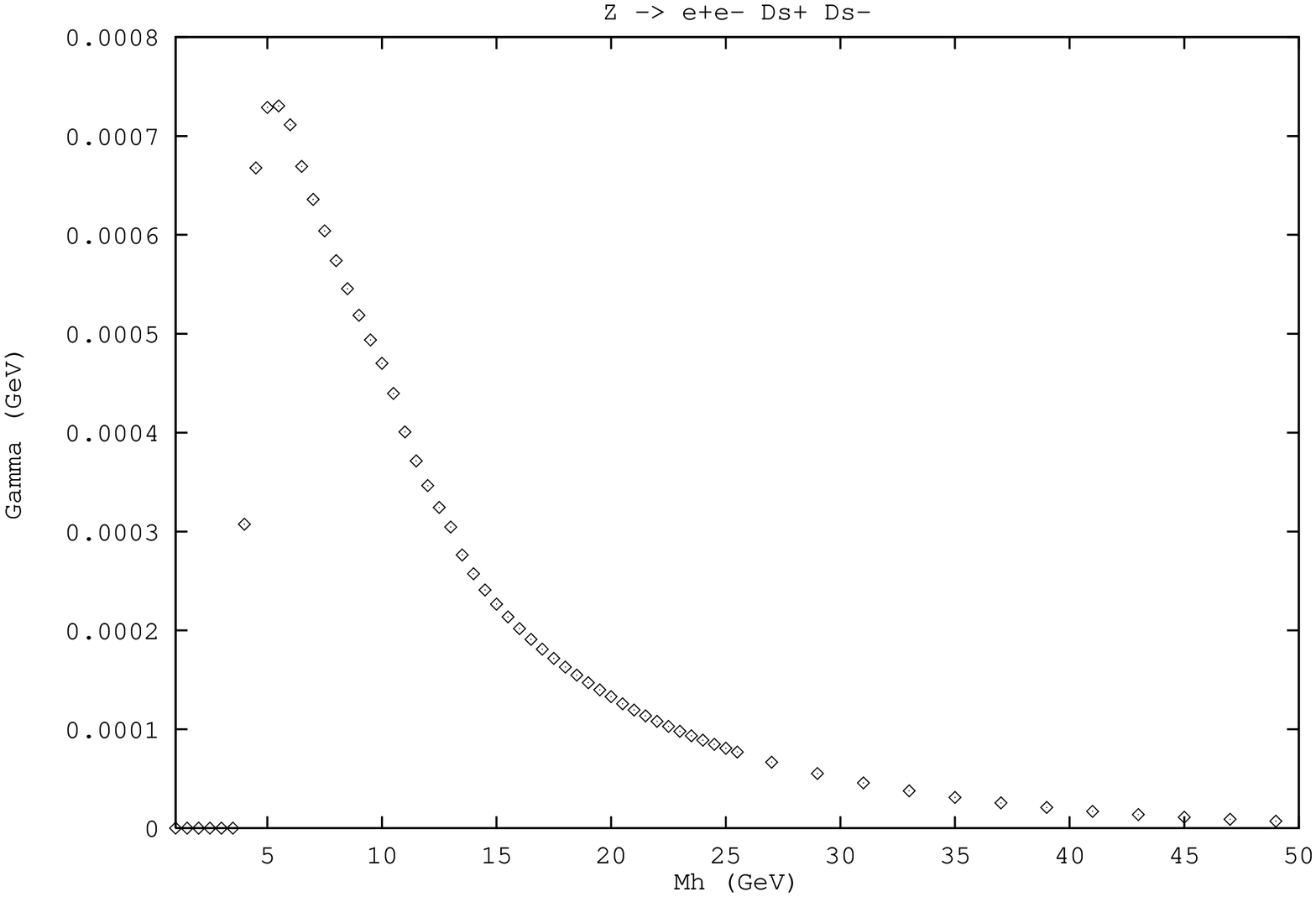,height=14truecm,width=10truecm,angle=-90}
\figskip
{\em Fig.~5a: The decay rate $Z \rar e^+e^- D_s^+D_s^-$.}
\ect
}

\vbox{
\figskip
\bct
\epsfig{file=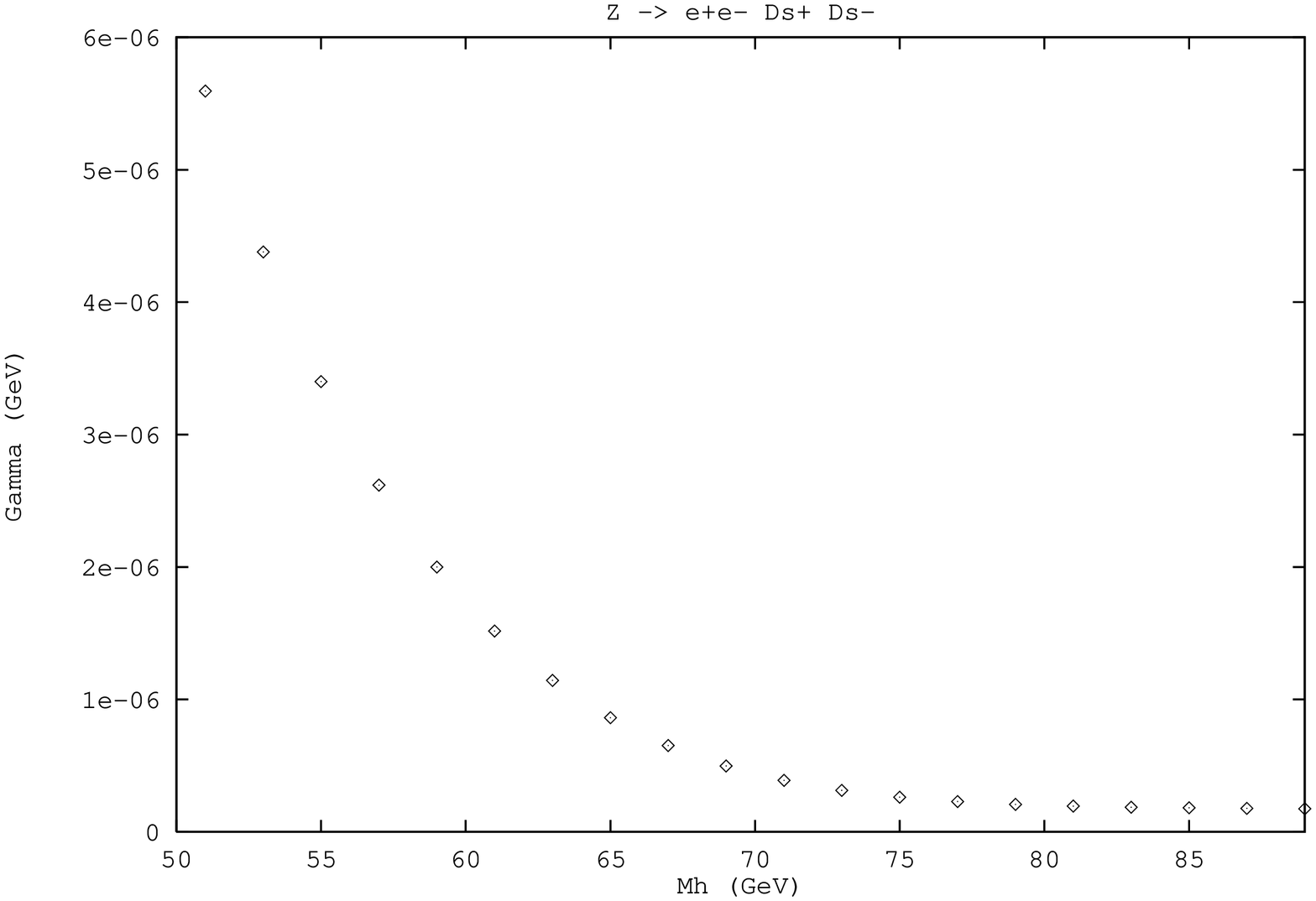,height=14truecm,width=10truecm,angle=-90}
\figskip
{\em Fig.~5b: The decay rate $Z \rar e^+e^- D_s^+D_s^-$.}
\ect
}

The threshold of observability
(\ref{eq:thresh}) is satisfied for $M_H \leq 65\,GeV$.

This is to be compared with the present lower bound of the four LEP
experiments for the Higgs boson of the standard model
$M_{H\,standard}^{LEP} \geq 90\,GeV$ at $95\%$ confidence level
\cite{HiggsLEP}.
In this mass range, there is no hope to detect the Higgs that is proposed
here, but it could have been missed at lower masses (but high enough
for the missing energy channel to be undetectable \cite{Higgsexp}) because
of a too low statistics.

\subsection{The decay $\boldsymbol{Z \rightarrow e^+e^-\,K^\pm\pi^\mp}$}
\label{subsec:KPI}

There also exist in this approach  decays which do not occur in the
Glashow-Salam-Weinberg model for quarks and characterize a composite Higgs
like has been introduced above; they are the ones
corresponding to ``flavour changing neutral currents'' in the scalar sector.
They could furthermore be easily identified experimentally if produced with
enough statistics.

{From} (\ref{eq:phi4}) and (\ref{eq:goldstone}) or its generalization to three
generations, it appears that the Higgs can decay into final states like
$K^\pm \pi^\mp, D^\pm K^\mp \ldots$, and that, due to the choice of the
kinetic terms as stated in subsections \ref{subsubsec:lagrangian} and 
\ref{subsubsec:invariants} and to the property of
diagonalization of the corresponding quadratic invariant
(\ref{eq:diaginvar}), the background described in the two previous decays is
now absent.

I study here the $e^+ e^- K^\pm \pi^\mp$ final state, the amplitude
of which benefits from a moderate damping by the mixing angles, only
$s_\theta c_\theta$.
The decay rate is computed form (\ref{eq:gammaBB},\ref{eq:largeur}) where
the factor ``$1$'' in each of them, corresponding to the direct coupling, is
dropped since the latter does not exist any more.

The results for the decay rate are plotted in Fig.~6.

\vbox{
\figskip
\bct
\epsfig{file=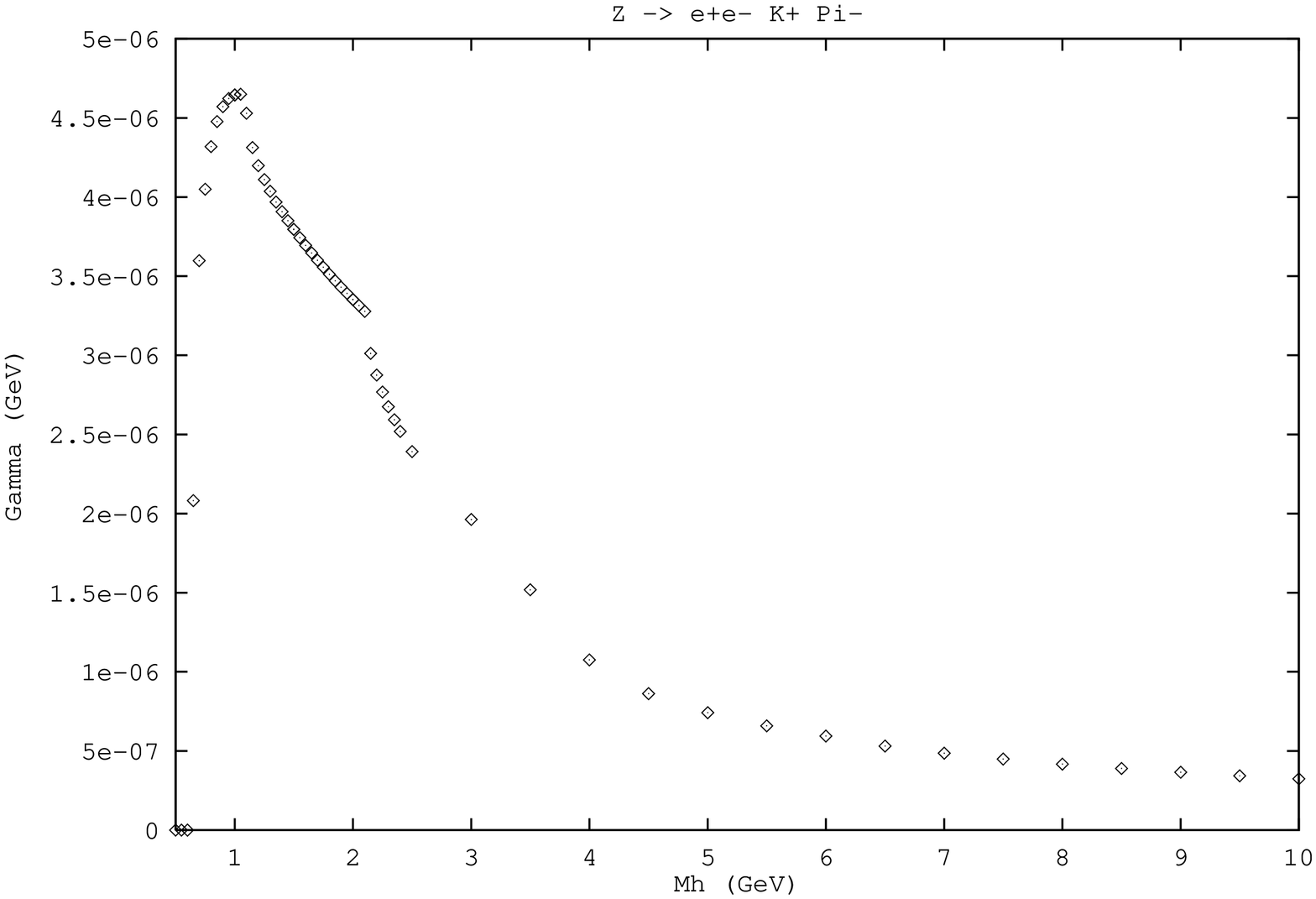,height=14truecm,width=10truecm,angle=-90}
\figskip
{\em Fig.~6: The decay rate $Z \rar e^+e^- K^\pm \pi^\mp$.}
\ect
}

It exhibits the same resonance-like behaviour as the previous decays, but has
its maximum $\Gamma_{max} \approx 4.7\,10^{-6}\,GeV$ at a low Higgs mass
$M_H \approx 1\,GeV$. The cusps that can be seen on the curve have been
checked to
correspond to the opening of the different two-mesons channels in the
propagator of the Higgs boson.

The resonance is rather sharp, and
it appears that for $M_H > 7\,GeV$ the width of the process is lower than
the threshold (\ref{eq:thresh}) above which there is no hope that it be
observed.

As a Higgs mass $M_H < 7\,GeV$ would have been detected by the corresponding
missing energy \cite{Higgsexp} (with the possible {\it caveat} of footnote
\ref{foot:asympt}), there seems unfortunately to be no hope 
to observe the characteristic decays mentioned above in a foreseeable
future.

\section{Conclusion}
\label{section:conclusion}

Answering a demand that alternatives to the strictly standard
Higgs boson or to its supersymmetric extensions be searched for.
I have proposed one, which, though it shares many similarities
with the standard model, exhibits a Higgs boson that is still more
elusive; it may not however be beyond experimental reach.

It  has been incorporated in a framework where the
fields in the Lagrangian are tightly related with the $J=0$ mesonic states
observed asymptotically; I have related the orientation of the
Higgs in flavour space to leptonic decay constants of pseudoscalar mesons
and to the CKM mixing angles, such that the quartic potential which triggers
the breaking of the electroweak symmetry has been expressed in terms of
these parameters and of the mass of the Higgs itself.
It includes in particular the coupling of the Higgs to two pseudoscalar
mesons, which is no longer triggered by Yukawa couplings to quarks followed
by an hadronization process as it used to be in the standard model.
New links have thus been provided, which enabled the study of the
disintegration of the $Z$ gauge boson into two leptons and two pseudoscalar
mesons.

I showed that this Higgs boson has interesting and specific properties, in
particular that it can trigger flavour changing neutral currents, by
decaying into final states of the type $K^\pm \pi^\mp$; unfortunately,
detecting those decays would require a tremendous increase of the available
number of $Z$ bosons.

The decay $Z \rar e^+e^- D_s^+ D_s^-$ is the best candidate because it is
not damped by small mixing angles and the background due to the direct
coupling of the $Z$ to two mesons is negligeable.
A Higgs with mass lower that $65\,GeV$ could have been missed.

Unlike in the standard model for quarks, the final state
$e^+e^- B^+B^-$ appears to be undetectable because of the presence of the CKM
mixing angles in the Higgs coupling to two pseudoscalar mesons; 
the Higgs contribution is furthermore screened by a background which,
though much larger, is itself undetectable.

I did not explicitly present here the results for another channel which is
not suppressed by small mixing angles, $Z \rar e^+ e^- \pi^+ \pi^-$:
the reason is that the corresponding decay rate is then peaked at very low
Higgs masses $M_H < 1\,GeV$, for which such a particle
could not have escaped detection in the missing energy channel; for higher
masses it becomes again absolutely undetectable.

For large Higgs masses ($M_H > M_W$), the type of decays studied here is
undetectable, and more standard considerations have to be pursued.

\vskip 1cm 
\begin{em}
\underline {Acknowledgments}: it is a pleasure to thank F. Boudjema for
critics and suggestions.
\end{em}
\newpage\null
\listoffigures
\bigskip
\begin{em}
Fig.~1:  Leptonic decay of a pseudoscalar meson;\l
Figs.~2:  The two contributions to the decay $Z \rightarrow e^+e^-\; B^+B^-$;\l
Fig.~3:  Introducing a width for the Higgs boson;\l
Fig.~4:  $10^4 \times$ the Higgs contribution to the rate of the decay
$\Gamma_{Z \rightarrow e^+e^-\, B^+B^-}$ as a function of the Higgs mass;\l
Figs.~5:  The decay rate $\Gamma_{Z \rightarrow e^+e^-\, D_s^+D_s^-}$ as a
function of the Higgs mass;\l
Fig.~6:  The decay rate $\Gamma_{Z \rightarrow e^+e^-\, \pi^+K^-}$ as a
function of the Higgs mass.
\end{em}

\newpage\null
{\Large\bf Appendix}
\appendix
\section{Diagonalizing eq.~(\protect\ref{eq:diaginvar}) in the basis of strong
eigenstates: a choice of $\mathbb D$ matrices}
\label{app:Dmatrix}

The property stated in subsection \ref{subsubsec:invariants}
 is most simply verified for the ``non-rotated''
$SU(2)_L \times U(1)$ group and representations which corresponds to
setting ${\mathbb K}={\mathbb I}$ in (\ref{eq:SU2L},\ref{eq:reps}).

\subsection{$\boldsymbol{N=2}$ (1 generation).}

Trivial case: $\mathbb D$ is a number.

\subsection{$\boldsymbol{N=4}$ (2 generations).}

The four $2\times 2$ $\mathbb D$ matrices ($3$ symmetric and $1$
antisymmetric) can be taken as
\begin{equation}
{\mathbb D}_1 = \left( \ba{cc} 1 & 0 \cr
                            0 & 1     \ea \right),\
{\mathbb D}_2 = \left( \ba{rr} 1 & 0 \cr
                            0 & -1    \ea \right),\
{\mathbb D}_3 = \left( \ba{cc} 0 & 1 \cr
                            1 & 0     \ea \right),\
{\mathbb D}_4 = \left( \ba{rr} 0 & 1 \cr
                           -1 & 0     \ea \right).
\end{equation}

\subsection{$\boldsymbol{N=6}$ (3 generations).}

The nine $3 \times 3$ $\mathbb D$ matrices ($6$ symmetric and $3$
antisymmetric),
can be taken as
\bea
&&\hskip -1cm {\mathbb D}_1 = \sqrt{{2\over 3}}\left( \ba{ccc}
                                1  &  0  &  0  \cr
                                0  &  1  &  0  \cr
                                0  &  0  &  1 \ea \right), \cr
&& \cr
&& \cr
&&\hskip -1cm {\mathbb D}_2 ={2\over\sqrt{3}} \left( \ba{ccc}
                \sin\alpha  &     0    &    0    \cr
       0     & \sin(\alpha\pm{2\pi\over 3})&   0   \cr
       0     &                    0        & \sin(\alpha\mp{2\pi\over 3})
       \ea\right),\
{\mathbb D}_3 ={2\over\sqrt{3}} \left( \ba{ccc}
                \cos\alpha  &     0    &    0    \cr
       0     & \cos(\alpha\pm{2\pi\over 3})&   0   \cr
       0     &                    0        & \cos(\alpha\mp{2\pi\over 3})
       \ea\right), \cr
&& \cr
&& \cr
&&\hskip -1cm  {\mathbb D}_4 =\left( \ba{ccc}
                                0  &  0  &  1 \cr
                                0  &  0  &  0 \cr
                                1  &  0  &  0   \ea \right),\
     {\mathbb D}_5 =\left( \ba{rrr}
                                0  &  0  &  1 \cr
                                0  &  0  &  0 \cr
                               -1  &  0  &  0   \ea \right),\cr
&& \cr
&& \cr
&&\hskip -1cm {\mathbb D}_6 = \left( \ba{ccc}
                                0  &  1  &  0  \cr
                                1  &  0  &  0  \cr
                                0  &  0  &  0   \ea \right), \
 {\mathbb D}_7 = \left( \ba{rrr}
                                0  &  1  &  0  \cr
                               -1  &  0  &  0  \cr
                                0  &  0  &  0   \ea \right), \
 {\mathbb D}_8 = \left( \ba{ccc}
                                0  &  0  &  0  \cr
                                0  &  0  &  1  \cr
                                0  &  1  &  0   \ea \right), \
 {\mathbb D}_9 = \left( \ba{rrr}
                                0  &  0  &  0  \cr
                                0  &  0  &  1  \cr
                                0  & -1  &  0   \ea \right),\cr
& &
\eea
where $\alpha$ is an arbitrary  phase.

\section{The normalization of the asymptotic mesons and the
Gell-Mann-Oakes-Renner relation}
\label{app:GMOR}
{From} (\ref{eq:flavoureigenstates}), (\ref{eq:bdef}) and (\ref{eq:b})
one gets for the pseudoscalar asymptotic meson ${\cal P}_{ab}$ associated
with the matrix ${\mathbb M}_{ab}$ (see subsection \ref{subsubsec:flavour})
\begin{equation}
{\cal P}_{ab}({\mathbb M}_{ab}) =
     i\,\frac{2f_0}{\la \ol\Psi {\mathbb H} \Psi\ra}
                 \ol\Psi \gamma_5 {\mathbb M}_{ab} \Psi,
\label{eq:meson}
\end{equation}
where the matrix $\mathbb H$ which describes the Higgs boson $H$ in
flavour space has been defined in (\ref{eq:Hnorm}).

To make the link with the customary quark picture, one recalls that, in QCD,
the pseudoscalar density $i \ol\Psi\gamma_5 {\mathbb M}_{ab} \Psi$ is
related to the divergence of the axial current with flavour indices $a,b$
\begin{equation}
A^\mu_{ab} = \ol \Psi \gamma^\mu\gamma_5 {\mathbb M}_{ab} \Psi
           = \bar q_a \gamma^\mu \gamma_5 q_b
\label{eq:current}
\end{equation}
by
\begin{equation}
\p_\mu A^\mu_{ab} = i(m_a + m_b) \ol\Psi\gamma_5 {\mathbb M}_{ab} \Psi,
\label{eq:divergence}
\end{equation}
where $m_a$ and $m_b$ are the ``masses'' of the quarks $q_a$ and $q_b$.

Now, one form of the PCAC statement is that $\p_\mu A^\mu_{ab}$ is
proportional to the interpolating field of the observed pseudoscalar
meson ${\cal P}_{ab}$ according to
\begin{equation}
\p_\mu A^\mu_{ab} = f_{ab} M^2_{ab} {\cal P}_{ab},
\label{eq:PCAC}
\end{equation}
where $M_{ab}$ is the mass of the meson.
This transforms (\ref{eq:meson}) into
\begin{equation}
{\cal P}_{ab}({\mathbb M}_{ab}) =
     \frac{2f_0}{\la \ol\Psi {\mathbb H} \Psi\ra}
     \frac{f_{ab} M^2_{ab}}{m_a + m_b}\  {\cal P}_{ab}
\end{equation}
which entails
\begin{equation}
(m_a + m_b) \la \ol\Psi {\mathbb H} \Psi\ra = 2\,f_0 f_{ab}\;M^2_{ab}.
\label{eq:GMOR}
\end{equation}
Eq.~(\ref{eq:GMOR}) is, up to a sign (unimportant since it can be introduced
in (\ref{eq:meson})), the equivalent of the Gell-Mann-Oakes-Renner relation
\cite{GellMannOakesRenner},
showing the consistency of our approach with a more traditional one.

%
\newpage\null
\begin{em}

\end{em}

\end{document}